A Physically-Intuitive Method for Calculation of the Local Lattice Constant from a High-Resolution Transmission Electron Microscopy Image by Fourier Analysis

James T. Teherani*, Judy L. Hoyt

Microsystems Technology Laboratories, Massachusetts Institute of Technology, Cambridge, MA 02139, USA

* Corresponding author. Tel.: +1 617 452 2873; *E-mail address*: teherani@mit.edu

**Abstract**

We have developed a physically-intuitive method to calculate the local lattice constant as a function of position in a high-resolution transmission electron microscopy image by performing a two-dimensional fast Fourier transform. We apply a Gaussian filter with appropriate spatial full-width-half-max (FWHM) bandwidth to the image centered at the desired location to calculate the *local* lattice constant (as opposed to the *average* lattice constant). Fourier analysis of the filtered image yields the vertical and horizontal lattice constants at this location. The process is repeated by stepping the Gaussian filter across the image to produce a set of local lattice constants in the vertical and horizontal direction as a function of position in the image. The method has been implemented in a freely available tool on *nanoHUB* [1].

1. Introduction

Strain engineering has been widely used to improve device performance as strain alters the band structure of materials which directly impacts band alignment, band gap, effective mass, and carrier mobility [2–5]. The strain state can be calculated as the ratio of the change in lattice constant to its unstrained value. Spatially resolved strain analysis of new device structures is necessary to understand electron and hole transport in these devices.

Several techniques that extract *local* lattice parameters (and therefore strain) from a device structure currently exist but require specialized measurements (convergent beam electron diffraction [6], dark field holography [7,8]), specific fabrication structures (nanobeam electron diffraction [9]), or complex mathematical analysis (geometric phase analysis [10]). Here, *local* refers to the lattice constant at a specific position as opposed to the average of the entire imaged structure. In this work, we develop a physically-intuitive method to extract the local lattice parameters from a conventional high-resolution transmission electron microscopy (HRTEM) image. The technique is especially useful in the analysis of nanostructure devices with non-uniform strain [11].

The developed method can be used in conjunction with any of the aforementioned methods to verify results from the various techniques, and it provides a physical framework from which to interpret the calculations. Additionally, we implement this method as a free online tool for the scientific community through the *nanoHUB* website [1] where the source code is also provided.

## 2. Background

A HRTEM of a crystalline zinc blende semiconductor (e.g., silicon, germanium, and gallium arsenide) resembles a fine checkerboard pattern with bright or dark spots at the location of atomic lattice sites depending on the type of imaging used. This checkerboard pattern can be approximated by the product of two sine functions such that

$$f(x,y) = \sin\left(\frac{2\pi}{\lambda_x}x\right)\sin\left(\frac{2\pi}{\lambda_y}y\right)$$

where $\lambda_x$ and $\lambda_y$ are the wavelengths of the pattern (i.e. lattice spacing) in the orthogonal $x$- and $y$-directions. The two-dimensional Fourier transform of such a function is given by

$$\mathcal{F}[f(x,y)](u,v) = \iint_{-\infty}^{\infty} e^{-2\pi i(ux+vy)} f(x,y)dxdy$$

In the equation, $u$ and $v$ are orthogonal reciprocal lattice vectors along the $x$- and $y$-directions. The Fourier transform solution for an infinite $f(x,y)$ is the sum of four Dirac-delta functions centered at $(u_o, v_o)$, $(-u_o, v_o)$, $(-u_o, -v_o)$, and $(u_o, -v_o)$ where $u_o = 1/\lambda_x$, and $v_o = 1/\lambda_y$. If the original checkerboard pattern is rotated by $\theta$ with respect to the $x$-axis, then the Dirac-delta function locations of the Fourier transform will be rotated by $\theta$ with respect to the corresponding $u$-axis, as shown in Fig. 1.

Conceptually, a Fourier transform maximum represents the strongest spatial frequency component of the original image. The experimental $f(x,y)$ will be noisy and finite, which will smear the Dirac-delta peaks of the Fourier transform. Additionally, spatial variation of the wavelength will cause variation of the spatial frequency, which will also smear the peaks. The net result is that the Fourier transform peak location represents the most dominant spatial frequency of the entire function though other spatial frequencies may also exist. The Fourier transform can be performed on a limited region of the original function to find the dominant spatial frequency of that region. In this work, we apply a Gaussian filter centered at a specific location to limit the sampled region of the original image.

The lattice spacing is equal to the reciprocal of the spatial frequency. The lattice constant can be found from the lattice spacing; for a zinc blende material with the $x$-axis oriented along the $[hkl]$ Miller direction, the lattice constant is given by

$$a_x = 2d_x\sqrt{h^2 + k^2 + l^2}$$

where $a_x$ is the lattice constant and $d_x$ is the lattice spacing in the $x$-direction. The equation is also valid for the $y$-direction with appropriate substitutions.

## 3. Methods

Several steps are required in the calculation of the local lattice constant as a function of position, and the procedure is presented in Fig. 1. We begin with the original HRTEM image and then apply a Gaussian filter centered at the desired position. The Gaussian filter has the effect of narrowing the area of the image over which the lattice constant is calculated. Next, we compute the two-dimensional fast Fourier transform (FFT) of the filtered image. We correct for rotation of the original image and find $u_o$ and $v_o$, the reciprocal lattice vectors in the horizontal and vertical directions, from the maxima of the Fourier transformed image. From the reciprocal lattice vectors, we calculate the lattice constant, and we repeat this procedure by scanning the Gaussian filter across the original image to obtain the lattice constant as a function of position in the original image.

The full-width-half-max (FWHM) of the Gaussian filter determines the area over which the lattice spacing is calculated from the HRTEM image, i.e. the number of lattice sites over which the FFT is calculated; the FFT of a Gaussian filtered image with a small FWHM samples only lattice sites near the position of interest yielding a *local* lattice constant whereas a large FWHM samples over many lattice sites and represents the *average* lattice constant over a large area.

Small rotations in the original image can cause an error in the peak positions of the FFT that are used to calculate the lattice spacing. The resulting error in the lattice spacing due to rotation may be of the same magnitude as the effect of strain. We correct for rotations in the original HRTEM image by using the method shown in Fig. 2.

## 4. Results

We applied the method to HRTEM images of a planar structure [12] and a trigate structure [11] shown in Figs. 3a and 3b. The planar structure is comprised of a relaxed $Si_{0.7}Ge_{0.3}$ virtual substrate on which a 5-nm strained Ge (s-Ge) layer and 5-nm strained Si (s-Si) layer is grown. As shown in Fig. 3c, the lattice constant in the $x$-direction ($a_x$) is maintained in all layers due to pseudomorphic growth conditions which results in compressive biaxial strain for the Ge layer and tensile biaxial strain for the Si layer. In-plane compressive biaxial strain results in tensile out-of-plane ($y$-directed) strain and vice versa. This principle is exemplified by the calculation of vertical lattice constant ($a_y$) shown in Fig. 3e; $a_y$ for s-Ge is larger than its relaxed value 5.658 Å, and $a_y$ for s-Si is smaller than its relaxed value of 5.431 Å.

The trigate structure (Fig. 3b) consists of a 8.5-nm s-Ge layer directly on high-$\kappa$ dielectric with a 4-nm Si capping layer with further fabrication details explained in Reference [11]. The Ge and Si layers are non-uniformly strained due to the nanoscale patterning that relaxes some of the strain in the horizontal direction. The analysis shows the highly non-uniform nature of the strain, and the calculated lattice constants match well with theoretical simulations (not shown) that seek to minimize elastic energy in the structure. The non-uniform strain causes significant changes in the energy bands of the material which greatly affect carrier localization, mobility, and device performance.

## 5. Conclusions

In this work, we describe and implement a method to calculate the horizontal and vertical lattice constants as a function of position from a HRTEM image. For the technique, we compute the two-dimensional

Fourier transform of a Gaussian filtered HRTEM image. From the location of peaks of the Fourier-transformed image, the local lattice constants in the vertical and horizontal direction are calculated. The method allows lattice constant extraction in non-planar nanostructures in which non-uniform strain occurs and can be used in conjunction with other methods, such as Raman spectroscopy, to determine the strain state of the material.

**Acknowledgments**

We would like to acknowledge W. Chern, P.Hashemi, and C. Ni Chleirigh for providing HRTEM images. In addition we would like to acknowledge S. Gradecak and D. Antoniadis for helpful discussions. J. Teherani acknowledges support from a U.S. Department of Defense NDSEG Fellowship. This work was supported in part by the NSF Center for Energy Efficient Electronics Science under NSF Award ECCS-0939514.

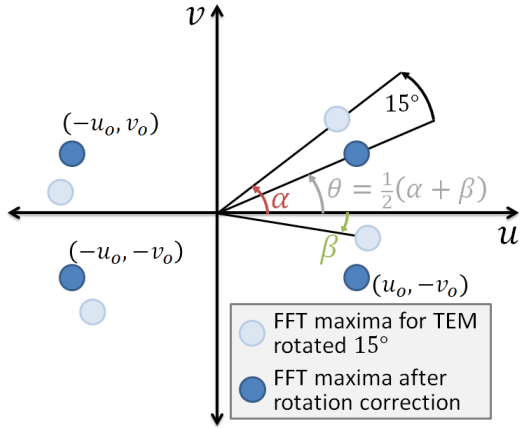

**Fig. 1.** Locations of the maxima of the Fourier transform of the Gaussian filtered HRTEM plotted as a function of reciprocal lattice vectors $u$ and $v$ corresponding to the $x$- and $y$-directions. The light blue circles correspond to maxima in which the lattice is rotated 15° with respect to the $x$-axis. The dark blue circles represent the position of the maxima after correction for rotation. The angle of the rotation-corrected maximum, $\theta$, is equal to the average of $\alpha$ and $\beta$, the angles of the rotated maxima.

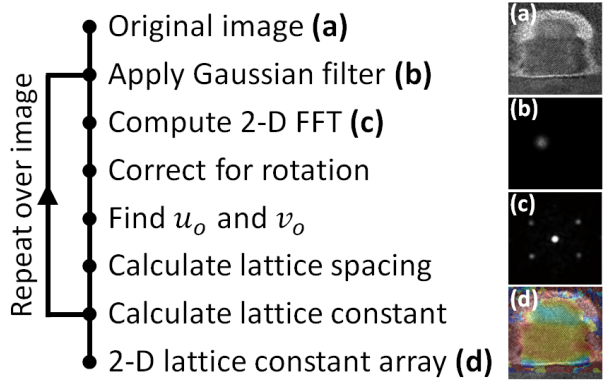

**Fig. 2.** Procedure used to calculate the lattice constant as a function of position of the original HRTEM image. (a) HRTEM image of a heterostructure tri-gate metal-oxide-semiconductor field-effect transistor (MOSFET). (b) Gaussian filter applied to original image. (c) Two-dimensional fast Fourier transform of (b). The center maximum represents the average pixel intensity of the transformed image and is not used in the calculation. The locations of the other four maxima are used to calculate the lattice spacing from which the lattice constant is computed. (d) The computed lattice constant in the vertical direction as a function of position overlaid on the original HRTEM image. HRTEM image courtesy of W. Chern and P. Hashemi [11].

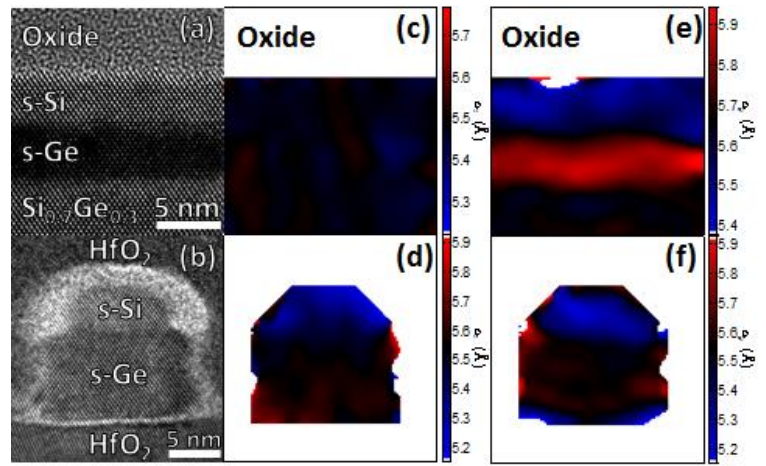

**Fig. 3.** HRTEM images of (a) planar [12] and (b) trigate s-Ge/s-Si heterostructures [11]. The planar structure is biaxially strained while the trigate structure has non-uniform strain due to the nanoscale patterning which relaxes some strain in the horizontal direction. The method to calculate local lattice constants is used to find the lattice constant in the horizontal direction (c-d) and the vertical direction (e-f) for both of the images.